\documentstyle[12pt,moriond]{article}
\begin{document}
\heading{Far Infrared Emission from Galaxy Clusters and Early-Type Galaxies}

\author{Joel N. Bregman$^{1}$, Caroline V. Cox $^{2}$} 
{$^{1}$ University of Michigan, Ann
Arbor, MI,
       USA.}  {$^{2}$ University of Virginia, Charlottesville, VA, USA.}

\begin{moriondabstract}

Infrared emission has been detected from normal elliptical galaxies and
from clusters of galaxies at 60$\mu m$ and 100$\mu m$ with the IRAS satellite.
In both cases, the emission has the characteristics of cool dust with
a temperature near 30K.  For the elliptical galaxies, there is a correlation
between the optical luminosities and the far infrared luminosities.  
The likely 
source of the dust in these systems
is mass loss from stars, which is heated by starlight and either is distributed
throughout the galaxy or falls into a central disk.  Analysis of upcoming 
ISO data will permit us to distinguish between these possibilities.

The far infrared emission from clusters of galaxies is of high luminosity
but is only detected in 10\% of the cases.  The heating is either due to
electron impact from the hot gas (a suitable explanation for one cluster) 
or photon absorption.  The source of the dust is probably gas stripped
from galaxies.  For clusters, ISO data will provide good mass determinations and
positions, thereby allowing us to determine the source of the heating.

\end{moriondabstract}

\section{Introduction}
Prior to the launch of the Infrared
Astronomical Satellite (IRAS),        
there were a variety of astronomical objects that were anticipated
to be sources of far infrared emission, such as gas-rich spiral galaxies and Galactic
star-forming regions.  Early-type galaxies and clusters of galaxies
were not among the objects that were expected to be detected with IRAS.  
Early-type galaxies have very little detectable cold gas from 21 cm studies
and are dominated by their hot X-ray emitting atmospheres \cite{Roberts}, 
while clusters of galaxies have vastly more substantial amounts of 
even hotter gas \cite{FNC}; in
these environments, dust would not survive long against sputtering by the 
hot gas nuclei.  Therefore, it came as a surprise when far
infrared emission was detected from these systems.

In elliptical galaxies without active galactic nuclei (AGN), the FIR emission 
indicates the presence of gas and dust shed during normal stellar evolution,
or perhaps captured during a merger event with a dwarf irregular.
In clusters of galaxies without AGN, the FIR emission is most likely due to 
gas and dust that has been stripped from spiral galaxies in the cluster, possibly in 
catastrophic fashion.
The study of FIR emission in these two different situations can broaden our 
understanding of these different systems.  Here we describe the present situation 
as determined from IRAS data and discuss the likely gains that data obtained with the 
Infrared Space Observatory (ISO) will provide.

\section{Far Infrared Emission From E and E/S0 Galaxies}
While early type galaxies have the same fraction of mass in gas as spiral galaxies,
the composition of their interstellar medium is strikingly different. 
Most of the the gas in early-type galaxies is hot ($10^{6}$K - $10^{7}$K) 
X-ray emitting 
material, and only a small amount is in a cool phase.  It is this cool phase 
that will be of concern in
this talk because it can be particularly revealing about the nature and
evolution of these galaxies and their stellar systems.
 
Cold material can be built-up in early-type galaxies either through the
capture of external gas or by accumulated stellar mass loss.  In the 
former case, a dwarf galaxy may collide with the larger elliptical so that 
its cold gas becomes part of the galaxy immediately.  Alternatively, a dwarf
galaxy may pass through the hot atmosphere of the early-type galaxy 
(or that of a group or cluster in which it resides) and have its cold material 
stripped away; eventually this material falls into the
elliptical, as appears to be the case for a dwarf galaxy UGC 7636 near 
NGC 4472 \cite{MSHJ}.  In such cases, the orbit of the captured
gas often will not be rotating in the same direction as the galaxy and the
axis of rotation may not be aligned with minor axis of the elliptical.  
In these cases, some gas can be identified as being captured, through 
careful 21 cm mapping.  If a considerable amount of gas is captured
(e.g., $10^{9}$ M$\odot$), these dramatic events may reshape the nature 
of the galaxy, leading to a disk of stars in addition to the spheroid.

A constant source of cold gas for elliptical galaxies is the mass lost from
stars, which accounts for 0.1-1 M$\odot$/yr in moderately bright elliptical 
galaxies.  The fate of this cold gas is not entirely clear because the 
interaction with the hot atmosphere has not been extensively studied.  Most 
of the mass
loss occurs during a brief evolutionary period near the red giant and 
planetary nebular phase of low mass stars (e.g., main sequence masses of 
0.8-1.0 M$\odot$).  
The mass shed from the star interacts with the hot atmosphere of the galaxy 
and is expected to be brought to rest by shocks; in the process, the shocks
heat the gas to the temperature of the hot galactic ISM.  The first
detailed calculations of this complicated process are now being carried
out \cite{P}.  These calculations indicate that a bow shock around
the star leads to a shock in the stellar mass loss that pushes the material
downstream but does not lead to a great deal of heating.  As this wake 
drifts downstream, there is a substantial velocity differential with the
hot atmosphere, so Kelvin-Helmholtz instabilities cause fingers of material
to be drawn out of the wake.  The mass in these fingers of material
is heated by shocks and  mixing,  while cooling occurs through optically thin 
radiative losses.  When the mass loss has its origin from a star that has a velocity 
significantly less than the velocity dispersion of the galaxy, the shock 
heating is reduced
and radiative losses are enhanced due to the tendency of material to seek
pressure equilibrium.  The calculations, now in progress, will
determine the fraction of mass lost from stars that remains cold and that 
which is converted into hot gas.

        Stellar mass loss that remains cold will sink toward the center of
the galaxy, and if it retains the net angular momentum of the stellar system,
it will settle into a small disk.  The size of the disk depends on where the 
gas was lost and the rotational velocity of the system, but the characteristic
size would be 1/4 - 1/30 the optical size of the stellar system.  The mass
of the disk of cold gas and dust would depend upon the the rate at which the
cold material is converted into stars and the fraction of the stellar mass
loss that remains cold.  Provided that the hot gas does not eventually 
evaporate the cold material, the dust in this cold gas can be long-lived 
because it is "protected" against sputtering by the hot gas.

        For the stellar mass loss that is heated to hot gas temperatures,
the fate of the dust is different.  The dust finds itself in a hot
environment and its drift time into the center of the galaxy is long, so
the dust interacts with the hot gas and photons locally, until it is 
destroyed by sputtering.  The sputtering time is approximately (e.g., \cite{Dwek})
\begin{equation}
7\times 10^{7}~~ ({n_{e}\over{3 \times 10^{-3} {\rm cm^{-3} } } })^{-1}~~ ({a \over {0.1 \mu m} }) {\rm yr}
\nonumber
\end{equation}
where $n_{e}$ is the electron density of the hot gas and $a$ is the grain size.
Sputtering is largely due to the protons and alpha particles colliding 
with the grains, while electron collisions lead to heating of the dust 
grains.  Ultraviolet and near-ultraviolet photons from elliptical
galaxies are another, probably more significant  source of heating of
the dust \cite{TM95}, \cite{TM96}.
 
        The dust that is mixed into the hot gas will be distributed on
a scale as large as the galaxy while the material that falls into the 
center of the galaxy will be relatively compact.  Which of these components
dominates the emission depends upon two factors: the rate of star
formation in the disk and the fraction of gas that remains cold and falls
into the disk.  This latter quantity can be calculated \cite{P},
and it may be possible to gain information about the star formation rate
through studies of the young stellar component (e.g., \cite{W}).
The dust properties can be studied directly from their thermal emission,
which has been possible through IRAS observations, and instruments on ISO
offer the possibility of much more detailed studies.  I will try to summarize
the state of the IRAS observations and indicate the potential of observations
being obtained with ISO.
 
        The IRAS satellite had four broad photometric bands centered at 12$\mu m$, 
25$\mu m$, 60$\mu m$, and 100$\mu m$, with the latter two being the most sensitive to the
thermal emission from dust mixed with cold gas in ordinary spiral galaxies.
Given the sensitivity of IRAS and the sensitivity of radio telescopes that
detect neutral hydrogen 21 cm emission and CO(1-0) 115 GHz emission, IRAS was capable of
detecting the presence of cold material that is below the detection threshold
of the radio telescopes (for typical dust to gas ratios in near-solar
abundance gas).  Consequently it was able to perform more sensitive searches 
for cold dust and gas in a variety of astronomical objects.  

        One class of astronomical object studied in this manner were 
early-type galaxies, where early studies indicated that the detection 
rate approached 50\% \cite{Jura}, \cite{Knapp}, \cite{Goudfrooij}.
This was a very exciting result as it indicated that dust was extremely
common in systems where cold gas was thought to be rare.
A particularly surprising result was that there was no clear
correlation between the apparent B magnitude and the FIR flux.
This would suggest that the intrinsic dispersion between the optical
and far infrared luminosities was quite large, another potentially important
result.
 
        As often occurs in detection studies, the significance of the FIR
emission in many of the galaxies was near the detection threshold, and we have 
learned considerably 
more about low signal-to-noise measurements since these initial studies on
the subject (see below).  The IRAS detection software provides a flux, uncertainty, and
position; at moderate or high S/N levels, these quantities are known
to be accurate.  However, at low S/N, the noise appears to be non-Gaussian
in nature, and studies of blank fields show that the spurious detection rate 
for sources near 3 sigma is 16\% at 60$\mu m$ and 31\% at 100$\mu m$ \cite{CBS}. 
This work indicates that sources above the 98\% confidence
threshold have S/N (from the standard SCANPI IRAS package) of 4 in the 60$\mu m$
filter and 4.5 at 100$\mu m$ .  The remainder of the study focuses on the 60$\mu m$
and 100$\mu m$
flux and the "bolometric" flux that is often formed from these two quantities:
\begin{equation}
F_{\rm FIR} = {1.257 \times 10^{-11}} ( {2.58F_{60} + F_{100}} )
\nonumber
\end{equation}
where the 60$\mu m$ and 100$\mu m$ fluxes are in Janskys.

        Given the above criteria for detection accuracy, we have reconsidered 
the rate of detection of early-type galaxies \cite{Bregman}.  The 
approach used here is different than that of Knapp \cite{Knapp}, which was a 
comprehensive survey that was intended to include all early-type galaxies,
including those that had nuclear activity.  Here we are asking whether 
normal non-interacting, non-peculiar early-type galaxies contain cold
material.  In doing so, we begin with all of the E and E/S0 galaxies
in the RSA catalog that are not listed as peculiar, do not possess nuclear
activity, are not near contaminating sources (e.g., spiral galaxies), and
do not lie in regions of rapidly changing Galactic cirrus emission.  After
excluding these sources, we find that 15 galaxies are detected above the
98\% confidence level (12\% of the sample) and an additional 7 galaxies are
detected in the 90-98\% confidence range (5\% of the sample).  This is a lower 
detection rate than found previously, but we do not mean to imply that previous works 
are wrong; the selection criteria imposed here are different and
quite strict.  Rather, the purpose of this work is to determine a sample
of "near-certain" detections to investigate whether these galaxies show a
connection with other galaxy properties.

        First, we compare the bolometric FIR flux to the optical B band
flux and we find a positive correlation, which indicates that optically 
bright objects are also brighter in the infrared.  Since we are only 
detecting the highest $F_{\rm FIR}$ objects, this study only defines the upper
envelope of the optical to FIR relationship, where we find that 
$F_{\rm FIR} \propto F_{\rm B}^{~0.24 \pm 0.08}$.  It is not particularly 
surprising that 
such a correlation exists, although it is relatively flat.  A positive 
correlation also exists between the FIR and blue luminosities, with
$L_{\rm FIR} \propto L_{\rm B}^{~1.65 \pm 0.28}$.  The temperature of the dust is in 
the range 23-38 K, with a median of 30K; the dust temperature is 
uncorrelated with $L_{\rm FIR}$.
 
        We have examined whether there are other signs of cold gas in these
FIR-detected galaxies and find that an unusually high fraction possess
other forms of cold or cool gas.  Of the sample of 15 excellent FIR detections,
four are detected in HI or CO, whereas only one detection was expected if
these galaxies were representative of the E and E/S0 galaxies in the RSA
catalog \cite{Roberts}.  Also, several of these galaxies have not
been observed at 21 cm or have very poor upper limits, so there are only a
few galaxies with stringent upper limits to their HI and CO content.  Since
less cold gas is required for a galaxy to be detected in the FIR than at
21 cm or in CO, the presence of FIR emission without HI or CO emission is 
not inconsistent.
Similarly, a larger fraction of these FIR galaxies have optical emission line
gas; 78\% were detected in $H_{\alpha}$, whereas only 29\% would 
be expected to show such lines in a randomly selected sample.  

          With this set of detections from non-AGN early-type galaxies, we
can investigate whether the dust is located in a central disk or distributed
throughout the galaxy.  This issue can be addressed by complementary observations 
as well
as by comparing models to the data.  Observationally, Goudfrooij \cite{Goudfrooij} has
obtained dust masses from extinction that can be compared to the dust masses
inferred from the FIR observations.  The extinction dust mass is usually
an order of magnitude smaller than the FIR dust mass (in our sample and also
in Goudfrooij's sample), which can be explained if most of the dust is
spatially distributed so that it does not create distinct dust lanes.
Given the uncertainties in determining dust masses from extinction 
observations and from FIR observations, we regard this as a tantalizing 
suggestion of distributed dust rather than definitive evidence;
it is interesting to note that most changes in our assumptions, such 
as the addition of cool dust, 
would increase the dust mass, indicating the existence of more distributed dust.

        The most detailed models yet developed are those by Tsai and Mathews
\cite{TM95}, \cite{TM96}, in which the dust from the stars is mixed with the hot gas where
it is slowly destroyed by sputtering while it is heated mainly by starlight.
Most of their models predict a luminosity that is too low and an effective
dust temperature ($F_{60\mu m}$ to $F_{100\mu m}$ ratio) that is too high relative
to the data.  However, 
for
the model where their maximum grain size is increased to 0.9$\mu m$, the 
$F_{60\mu m}$ to $F_{100\mu m}$ ratio
is similar to that observed, and the predicted infrared 
luminosity ($log(L_{\rm FIR}) = 42.66$) is comparable to that observed for most 
galaxies.
The galaxy with the highest $L_{\rm FIR}$ is NGC 7196, which is 6 times more
luminous than
the Tsai and Mathews predictions, and this is the only galaxy with CO emission.
It is likely that this galaxy has a central disk where most of the cold
material resides.  Without further detailed studies, it is difficult to 
determine how many of the galaxies are dominated by central cold disks, but
for most galaxies, the models can be explained with the infrared luminosity
being produced by distributed emission (e.g., mass loss from stars distributed
throughout the galaxy).

        In summary, one of the challenges in understanding the cold
interstellar medium in elliptical galaxies is to separate galaxies dominated
by captured gas from those where the gas is generated internally.  We have
concentrated on galaxies where the cold gas content has the greatest likelihood
of being generated internally by selecting non-peculiar E and E/S0 galaxies.
The IRAS emission is clearly detected in a modest fraction (12\%) of the 
galaxies and there is a correlation between the FIR flux and optical flux,
as well as between the FIR and optical luminosities.  For most of the galaxies,
the FIR emission is consistent with distributed emission, as would be expected
if we are detecting the emission of mass loss from stars and the mass loss is
distributed through the galaxy.  ISO holds the possibility of determining
directly whether whether or not the dust is distributed throughout the galaxy.
ISOPHOT has adequate resolution to map the shape of the emission in the 
nearest elliptical galaxies at 60$\mu m$-180$\mu m$ while ISOCAM can study the 
polycyclic aromatic hydrocarbon (PAH)
emission on angular scales of several arcseconds.  The data are now
being accumulated and analyzed and the results emerging in the next few years
should be very exciting.

\section{Far Infrared Emission From Clusters of Galaxies}

        The original motivation in searching for infrared emission from 
clusters of galaxies grew out of issues surrounding cooling flows in clusters.
The logic was that if gas is cooling at a rate of 100 M$\odot$/yr and forming
into stars at the same rate, that the star formation process may be 
accompanied by dust and infrared emission.  If the infrared signature were
scaled from spiral galaxies by the star formation rate, the emission would
be visible. The star formation rate in typical spiral galaxies is a few
solar masses per year, so a cooling flow cluster would be about 30 times 
more luminous in the FIR than an ordinary spiral galaxy.  Furthermore,
the emission would be expected from the cluster center where spiral galaxies
are rare and where the weakly emitting ellipticals would be undetectable at
the distances of most Abell clusters, so 
contaminating emission from galaxies would not occur.  Thus, the interpretation
of an infrared signature, should one be present, would be straightforward.
 
Two efforts searched for such emission in rich clusters, primarily utilizing
Abell clusters
at moderate redshifts (z = 0.02-0.2) and with strong known X-ray emission.
These investigators (\cite{BMO},\cite{GM})
reported that about 18-46\% of their samples of a few dozen 
clusters showed 60$\mu m$ or/and 100$\mu m$ emission above the 3$\sigma$ level;
nearly all of the detections were between the 3$\sigma$ and 5$\sigma$ level.  
There was no apparent correlation between the FIR properties and
the X-ray or optical properties
of the clusters.  The "detections" did not appear very convincing
to the eye, so we began a new and more thorough effort to understand the
noise properties and to enlarge the cluster sample size so that statistical studies
would become possible.
 
It is worth trying to understand what goes into the detection of an IRAS
source to understand the noise properties and contamination issues. IRAS 
was usually used in scanning mode whereby rectangular detectors swept over
a single point many times, usually with similar position angles.
At 60$\mu m$ and 100$\mu m$, the detector width is about 4.5' (by 1'), so the 
location information for
weak sources is generally quite poor, especially perpendicular to the
scan line.  Also, in crowded fields, the
width of the detector makes it quite easy for contamination to occur 
from more strongly emitting sources, such as a foreground spiral galaxy 
near the intended target.

However, perhaps the greatest problem with detecting sources is in
quantifying the statistical properties of the "baseline".  That is, 
it was hoped that as IRAS scanned across sources, one would see "baseline"
emission fluctuate about some low value with
a Gaussian distribution, with the source visible in clear contrast.  The
fluctuations are a combination of the Galactic Cirrus emission and the 
detector noise.  Unfortunately, we find that the fluctuations do not appear 
to have a normal distribution, and this has profound effects on the 
detection of weak sources \cite{CBS}.  We selected
about 200 blank fields and extracted a flux in the usual fashion,
using SCANPI (originally, the IRAS data product ADDSCAN).  We plotted a 
distribution of the
extracted "signals" and found it to be much broader than a Gaussian 
distribution.  We found that 16\% of the locations at 60$\mu m$ and
31\% at 100$\mu m$ had positive signals
at or above 3$\sigma$.  A simple criteria for
the true confidence as a function of SCANPI S/N was derived, and it
was found that for 98\%
confidence detections, one should use 4$\sigma$ at 60$\mu m$ and 4.5$\sigma$ 
at 100$\mu m$.
Note that 98\% confidence would correspond to +2$\sigma$ in Gaussian statistics.
The IPAC Faint Source Catalog is above these criteria
and does not suffer from a substantial number of spurious detections.
 
The sample of clusters used for this study included all of the Abell clusters with a dominant central galaxy 
(Bautz-Morgan class of I or I-II; \cite{Abell}) plus 
a few clusters with moderately high X-ray luminosities.  This led to 158
clusters for which IRAS data were available with a mean redshift of 0.076, 
and which forms a complete sample to a redshift of about 0.2.  We found that
10\% of the sample was detected above the 98\% confidence limit (18 clusters), 
which is significantly lower than previous studies.  The higher detection rates
of previous studies is partly due to spurious detections near the 3$\sigma$
level, but another factor is that the samples were chosen differently and
the mean distances were often different (e.g., the mean redshift was only
0.043 for the sample of Bregman, McNamara, and O'Connell \cite{BMO}).  
 
The spectrum of most detected clusters, as defined by their 60$\mu m$ and 100$\mu m$
fluxes and upper limits at 12$\mu m$ and 25$\mu m$, is similar to that expected
from cool dust and easily distinguished from starburst galaxies and AGNs.
This indicates that we are usually detecting emission from warm dust with
a typical temperature in the range 24-33 K.  No correlation was found with
cluster properties, although the detected clusters are slightly closer than
the sample as a whole.
 
Some of the scientific issues that are raised by this discovery are the
source of the dust and the origin of the heating for the dust.  The latter
issue is particularly acute because of the large FIR luminosities.  A
typical FIR luminosity is $10^{44.5}$ erg/sec, which is an order of magnitude
larger than the X-ray emission from within the core (cooling flow region; 
r = 100 kpc) and is comparable to the total X-ray luminosity of the cluster.
The FIR luminosity is typically a factor of 5 smaller than the bolometric 
optical emission from the central dominant galaxy.  Given these relative
luminosities, if the energy source is electron collisions by the hot gas,
then the cooling rate for the X-ray gas is an order of magnitude greater
than is usually assumed (e.g., the cooling rate is closer to 1000 M$\odot$/yr
rather than 100 M$\odot$/yr).  Since only 10\% of clusters are detected, 
the duty cycle of this intense cooling may be relatively brief.  If photons
heat the dust and the dust is superimposed upon the central dominant
galaxy, substantial reddening would be present.  Reddening toward these 
central galaxies has been searched for, but it is rarely detected and never
at the level needed to power the dust.  This does not rule out photon 
heating since the dust could be offset from the central galaxy, given the
poor IRAS resolution.  We note that in the nearby Centaurus cluster,
dust emission and extinction is observed in NGC 4696 \cite{Sparks},
but our clusters tend to be much more distant and the energy considerations
are more severe.
 
Heating by fast electrons makes predictions about the density of the hot gas,
and it is possible to to test these predictions through the use of
X-ray imaging data to determine the electron density in a given cluster.
The rate of electron collisions must be adequate to raise the dust temperature
to the level implied by the FIR data. ROSAT data was used to check this
for 5 clusters detected with IRAS.  We found that in Abell 1991, the electron 
density is above the value
needed to power the dust emission, but that for four other clusters (Abell 1541,
Abell 1691, Abell 2199 and Abell 2634), the electron density is too low by a 
substantial amount \cite{Cox95} \cite{Cox98}.
 
The dust mass is $10^{7}-10^{8}$ M$\odot$, so the associated amount of gas is 
$10^{9}-10^{10}$ M$\odot$, which is a typical mass for the interstellar medium of
a galaxy.  It is possible that as a galaxy passed through the core of
the cluster, gas and dust were stripped out and are now radiating.  The
sputtering time is short for dust in this environment, typically $10^{8}$ yr
in the cluster core ($n_{e} = 3 \times 10^{-3} cm^{-3}$, dust radius of 0.2 $\mu$m), so 
events
like this would need to be common unless the dust were protected by being
embedded in cooler gas.

It is hoped that ISO will be able to answer two of the primary issues in
this area:  the location of the dust and the dust mass.  The IRAS data showed
us that
there is dust emission somewhere within a $1'\times 4.5' $ box, but ISO will 
be able to locate the emission to a fraction of an arcminute, which should tell us 
if the dust is indeed cospatial with the dominant galaxy. ISOPHOT
maps of clusters detected with IRAS are being obtained for this project, 
and other teams have ISOCAM data of clusters, which will be valuable if there is 
significant PAH emission from
the dust.  IRAS also allowed us to estimate the mass of dust, but it simply
placed a lower limit on the dust mass since IRAS was not particularly sensitive
to
low temperature dust.  ISOPHOT has the capability to detect low temperature dust 
due to spectral windows that extend to longer wavelengths than IRAS; we
have utilized ISOPHOT's 180 $\mu m$ filter in our raster scans.
Although it is too early to report
our results, clusters have been detected by ISOPHOT and we hope that improved
data processing will lead to several interesting results.

\vfill
\end{document}